\documentclass[conference]{IEEEtran}

\usepackage{amsmath}
\usepackage{amsfonts}
\usepackage{graphicx}
\usepackage[short]{optidef}
\usepackage{newtxtext,newtxmath} % This was changed to help condense the equations
\usepackage{mleftright} % See command below
\mleftright % Reduce the space around \left and \right curly brackets
\usepackage[margin=0.70in]{geometry}
\usepackage{cite}
\newcommand{\Fourier}{\mbox{\boldmath$\cal F$}}

\IEEEoverridecommandlockouts
\begin{document}

\title{Efficient Precoding for Single Carrier Modulation in Multi-User Massive MIMO Networks}
\date{10 October 2020}
 \author{%
    \IEEEauthorblockN{Brent A. Kenney\IEEEauthorrefmark{1}, Arslan J. Majid\IEEEauthorrefmark{2}, Hussein Moradi\IEEEauthorrefmark{2}, and Behrouz Farhang-Boroujeny\IEEEauthorrefmark{1}}
    \IEEEauthorblockA{\IEEEauthorrefmark{1}Electrical and Computer Engineering Department, University of Utah, Salt Lake City, Utah, USA}
    \IEEEauthorblockA{\IEEEauthorrefmark{2}Idaho National Laboratory, Salt Lake City, Utah, USA\\}
\thanks{This manuscript has in part been authored by Battelle Energy Alliance, LLC under Contract No. DE-AC07-05ID14517 with the U.S. Department of Energy. The United States Government retains and the publisher, by accepting the paper for publication, acknowledges that the United States Government retains a nonexclusive, paid-up, irrevocable, world-wide license to publish or reproduce the published form of this manuscript, or allow others to do so, for United States Government purposes. \bf{STI Number: INL/CON-20-60127}.}
}

\maketitle
\begin{abstract}
By processing in the frequency domain (FD), massive MIMO systems can approach the theoretical per-user capacity using a single carrier modulation (SCM) waveform with a cyclic prefix.  Minimum mean squared error (MMSE) detection and zero forcing (ZF) precoding have been shown to effectively cancel multi-user interference while compensating for inter-symbol interference.  In this paper, we present a modified downlink precoding approach in the FD based on regularized zero forcing (RZF), which reuses the matrix inverses calculated as part of the FD MMSE uplink detection.  By reusing these calculations, the computational complexity of the RZF precoder is drastically lowered, compared to the ZF precoder.  Introduction of the regularization in RZF leads to a bias in the detected data symbols at the user terminals. We show this bias can be removed  by incorporating a scaling factor at the receiver.  Furthermore, it is noted that user powers have to be optimized to strike a balance between noise and interference seen at each user terminal.  The resulting performance of the RZF precoder exceeds that of the ZF precoder for low and moderate input signal-to-noise ratio (SNR) conditions, and performance is equal for high input SNR.  These results are established and confirmed by analysis and simulation.
\end{abstract}

%-------------------------------------------%
%-- Section I: Introduction --%
%-------------------------------------------%
\section{Introduction}
With the advent of digital antenna arrays, massive multiple input multiple output (MIMO) wireless communications system have become a reality.  Multiple user equipments (UEs) can operate on the same frequencies through spatial multiplexing techniques achieved through multi-user detection on the uplink (UL) and multi-user precoding on the downlink (DL).  A previous work has shown that when the number of base station (BS) antennas is much larger than the number of UEs, detection and precoding are successfully implemented using linear processing techniques \cite{Hoydis:2013}.  It was also noted in \cite{Hoydis:2013} that minimum mean squared error (MMSE) detection more rapidly approaches the massive MIMO UL capacity bound as the number of BS antennas increases, compared to matched filter (MF) detection.  Similar results were shown in \cite{Hoydis:2013} for zero-forcing (ZF) precoding compared to time-reversal (TR) precoding for the DL.

Massive MIMO networks with widely dispersive channels have traditionally used orthogonal frequency division multiplexing (OFDM) modulation.  OFDM is particulary attractive in these environments since it eliminates inter-symbol interference (ISI) by dividing up the signal into a large number of subcarriers.  Despite OFDM's popularity in current wireless standards, this modulation scheme has some drawbacks and is not optimized for every scenario.  For example, OFDM is often criticized for its high peak-to-average power ratio (PAPR), which results in a high power back-off to maintain linearity.  Consequently, OFDM systems have low power efficiency \cite{Tse:2005}, which is particularly detrimental to battery operated UEs.  It also affects DL operations at the BS.  

To alleviate the power efficiency problem with OFDM, single carrier modulation (SCM) has been reintroduced in a number of scenarios \cite{Benvenuto:2010}.  When framed with a cyclic prefix (CP) similar to OFDM, SCM can be effectively processed in a massive MIMO scenario using frequency domain (FD) processing, achieving the same capacity as OFDM.  While MF detection and TR precoding have been shown to be optimal for SCM for operation at low input signal-to-noise ratio (SNR) \cite{Ngo:2013} \& \cite{Larsson:2012}, multi-user interference (MUI) cannot be overcome at moderate to high input SNR unless SCM uses techniques such as MMSE detection \cite{Kenney:Globecom:2020} or ZF precoding \cite{Dinis:2018}.  The details of ZF precoding in the FD were presented in \cite{Dinis:2018}.  The implementation of the ZF solution is efficient given that the computational complexity, which is dominated by a matix inversion calculation, scales with the number of UEs rather than by the number of BS antennas.  

This paper provides the details for a regularized zero-forcing (RZF) precoder in the FD that reuses the MMSE-based matrix inverses that are calculated for the UL detector in \cite{Kenney:Globecom:2020}.  Because of the regularization factor in the matrix inverses, there are three special accommodations that are unique to the RZF solution.  First, a RZF-specific scale factor must be calculated to maintain the desired power at the BS, depending on the SNR operating point.  Second, on the UE detection side, another specialized scale factor is calculated to ensure an unbiased estimate.  Both of these scale factors can be pre-computed and accessed through table look-up.  The third accommodation results from the fact that RZF precoding does not perfectly cancel MUI, which becomes an issue when there are large variations in large-scale path loss for different UEs.  This issue is solved by optimizing the transmit power to each UE such that the output signal-to-interference-plus-noise ratio (SINR) is equal for all UEs.  The resulting performance of the RZF precoder exceeds that of the ZF approach for low and moderate input SNR and has the same performance at high SNR.  The improved performance of the RZF precoder may seem counterintuitive, but it will be shown that when the regularization term starts to dominate the matrix inverse, the precoder converges to TR precoding, which is optimal at low input SNR.  These results are accentuated by the fact that the computational complexity required for the precoder is reduced by an order of magnitude or more compared to the ZF precoder in \cite{Dinis:2018} for relevant numbers of simultaneous UEs.

The remainder of this paper is organized as follows: Section \ref{System_Model} introduces the system model that will be used to discuss DL precoding in the FD; Section \ref{Downlink_Precoding} details the calculations for precoding the multi-user signal at the BS and detecting the signal at each UE; Section \ref{Complexity_Comparison} compares the computational complexity of RZF precoding to ZF precoding; the RZF performance is analyzed in Section \ref{Performance_Analysis}; Section \ref{Simulation_Results} presents the results of the DL simulation to verify the analysis; and concluding remarks are in Section \ref{Conclusion}.

%-------------------------------------------%
%-- Section II: System Model --%
%-------------------------------------------%
\section{System Model} \label{System_Model}
The time domain duplex (TDD) scenario modeled in this paper assumes that the UEs do not have any channel state information (CSI).  Each UE is assumed to have knowledge of the cell parameters such as the number of BS antennas and number of simultaneous users.  The BS has the CSI between each UE and each of the $M$ BS antennas, which can be obtained by transmitting pilot signals from the UEs.  This paper assumes perfect CSI at the BS.  The effects of channel estimation are left to a future work.

The DL transmission is divided into frames, where the BS transmits $N$ unit variance symbols for each user.  The symbols designated for the $k^{\textrm{th}}$ UE are represented by the vector $\mathbf{s}_k$.  After precoding, a CP is added to the front of each frame to preserve circular convolution.  Since each user's frame is transmitted simultaneously with the other users, the frames must be precoded to limit MUI.  The precoding generates a unique transmit vector for each of the BS antennas.  The received DL signal for user $k$ after CP removal is expressed as
\begin{equation}
    \mathbf{r}_k = \frac{ 1 }{ \sqrt{p_k} } \sum_{ m=1}^M \mathbf{H}_{m,k} \mathbf{x}_m  + \mathbf{w}_k, 
    \label{eq:01}
\end{equation}
where $1 / \sqrt{p_k}$ is the scalar path loss between the BS to user $k$, $\mathbf{H}_{m,k}$ is the $N \times N$ circulant convolutional channel matrix between antenna $m$ and user $k$, $\mathbf{x}_m$ is the precoded signal vector for antenna $m$, and $\mathbf{w}_k$ is the receiver noise vector at the UE with variance $\sigma^2_w$.   Let $\mathbf{h}_{m,k}$ represent the channel impulse response vector between antenna $m$ and user $k$, which is of length $L_h$.  $\mathbf{H}_{m,k}$ is formed by taking $\mathbf{h}_{m,k}$, appending $N-L_h$ zeros to form ${\mathbf{h}_{m,k} }_{(0)}$, and then taking downward cyclic shifts of ${ \mathbf{h}_{m,k} }_{(0)}$ to create
\begin{equation}
  \mathbf{H}_{m,k}=
  \begingroup % keep the change local
    \setlength\arraycolsep{4pt}
    \begin{bmatrix}
       { \mathbf{h}_{m,k} }_{(0)} & { \mathbf{h}_{m,k} }_{(1)} \dots & { \mathbf{h}_{m,k} }_{(N-2)} & { \mathbf{h}_{m,k} }_{(N-1)}
    \end{bmatrix}
  \endgroup, 
    \label{eq:02}
\end{equation}
where the parenthetical subscript represents the number of downward cyclical shifts applied to the base vector.

For convenience and without loss of generality, the average channel power for each user over all BS antennas, $\frac{1} {M} \sum_{m=1}^M \mathbf{h}_{m,k}^{\textrm{H}} \mathbf{h}_{m,k}$, is normalized to unity.  Consequently, the input SNR is ${ 1 }/ {\sigma^2_w}$.  Note that the individual channel power values between each of the $M$ antennas and user $k$ are allowed to vary widely.

The approach taken in this paper is to perform precoding in the FD.  Taking the $N$-point Discrete Fourier Transform (DFT) of \eqref{eq:01} results in
\begin{equation}
    \tilde{ \mathbf{r} }_k = \frac{ 1 }{ \sqrt{p_k} } \sum_{m=1}^M \mathbf{\Lambda}_{m,k} \tilde{ \mathbf{x} }_m + \tilde{ \mathbf{w} }_k, 
    \label{eq:03}
\end{equation}
where the tilde represents the FD representation of the vectors.  The diagonal matrix $\mathbf{\Lambda}_{m,k}$ contains the eigenvalues of $\mathbf{H}_{m,k}$.  This results from the fact that the circulant matrix $\mathbf{H}_{m,k}$ is diagonalized by the DFT matrix,  $\Fourier$, where $\Fourier$ is scaled such that $\Fourier^{-1} = \Fourier^{\textrm{H}}$ (i.e., $\mathbf{H}_{m,k} = \Fourier^{-1} \mathbf{\Lambda}_{m,k} \Fourier$) \cite{Farhang:Adaptive_Filters}.  It follows that taking the $N$-point DFT results in $\Fourier \mathbf{H}_{m,k} = \mathbf{\Lambda}_{m,k} \Fourier$.  Let $\lambda_{m,k,i}$ represent the $i^{\textrm{th}}$ value along the diagonal of $\mathbf{\Lambda}_{m,k}$.  The eigenvalues can be obtained by taking the $N$-point DFT of the channel impulse response $\mathbf{h}_{m,k}$, which is used to form $\mathbf{H}_{m,k}$.  For more efficient computation, it is noted that all of the FD conversions can be performed with the Fast Fourier Transform (FFT) instead of the DFT.

The precoding is performed on a frequency bin basis.  We represent the $n^{\textrm{th}}$ bin of the received signal for each user as
\begin{equation} \small
	\begin{bmatrix}
    	  \tilde{r}_{1,n} \\
	  \tilde{r}_{2,n} \\
	  \vdots \\
	  \tilde{r}_{K,n}
	\end{bmatrix} = \mathbf{P}^{ - \frac{1}{2}}
	\begin{bmatrix}
    	  \lambda_{1,1,n} & \lambda_{2,1,n} & \dots & \lambda_{M,1,n} \  \\
    	  \lambda_{2,2,n} & \lambda_{2,2,n} & \dots & \lambda_{M,2,n} \  \\
	  \vdots \\
    	  \lambda_{1,K,n} & \lambda_{2,K,n} & \dots & \lambda_{M,K,n} \  \\
	\end{bmatrix} 
	\begin{bmatrix}
    	  \tilde{x}_{1,n} \\
	  \tilde{x}_{2,n} \\
	  \vdots \\
	  \tilde{x}_{M,n}
	\end{bmatrix} +
	\begin{bmatrix}
    	  \tilde{w}_{1,n} \\
	  \tilde{w}_{2,n} \\
	  \vdots \\
	  \tilde{w}_{K,n}
	\end{bmatrix}, 
    \label{eq:04}
\end{equation}
where $\mathbf{P}$ is a $K \times K$ diagonal matrix with values of $p_k$ and $ \tilde{x}_{m,n}$ is the precoded value in the FD for antenna $m$ and bin $n$.  This can be succinctly represented as
\begin{equation}
    \tilde{ \mathbf{r} }_{:,n} =  \mathbf{P}^{-\frac{1}{2}} \mathbf{A}_n^{\textrm{T}} \tilde{ \mathbf{x} }_{:,n} + \tilde{ \mathbf{w} }_{:,n}.
    \label{eq:05}
\end{equation}

%---------------------------------------------------------%
%-- Section IV: Downlink Precoding and Detection--%
%---------------------------------------------------------%
\section{Downlink Precoding and Detection} \label{Downlink_Precoding}
DL precoding is effectively performed in the FD for SCM waveforms with a CP in a massive MIMO scenario.  By processing in the FD, the precoding can simultaneously cancel the interference and pre-compensate for ISI.  A detailed analysis of DL precoding for SCM with a CP has been presented in \cite{Dinis:2018}, where the zero-forcing (ZF) algorithm is used to calculate the precoded vectors for each antenna.  The complexity of the ZF solution is mainly driven by a matrix inverse with dimension $K \times K$.  In this section, we take a different approach to the precoding, which is based on reusing the matrix inverse that is calculated for the UL detection as presented in \cite{Kenney:Globecom:2020}.  This solution drastically reduces the number of computations for the DL, compared to ZF precoding, without sacrificing performance.  Specifically, performance is maintained for high SNR operation and exceeds the ZF performance in the low SNR regime.

One aspect to consider for DL precoding is that the distance from the BS may vary considerably from user to user, resulting in a wide range of large-scale path loss values.  Traditionally, the BS transmits the signal intended for each UE at a power corresponding to the inverse of the large-scale path loss in order to achieve the same SNR at each UE.  This power variation is not a consideration for UL detection, where power control is assumed.  However, we will show that the matrix inverse calculated for the UL under power control conditions is still applicable to DL precoding, and performance can be optimized for large variations in the large-scale path loss by carefully allocating the power for each UE.

%-------------------------------------------------------------------%
%-- Subsection IV.A: ZF Precoding
%-------------------------------------------------------------------%
\subsection{Zero-Forcing Precoding}
The expression in \eqref{eq:05} shows the FD representation of the $n^{\textrm{th}}$ bin of the received signal vector for each of the UEs.  Each UE applies a scale factor to the received signal to obtain the FD symbol estimate (i.e., $\hat{ \mathbf{s} }_k = \mathbf{r}_k / \beta_{\textrm{ZF}}$ in this case).  We can express the signal in the FD without the addition of noise as
\begin{equation}
    \tilde{ \mathbf{s} }_{:,n} = \frac{1}{ \beta_{\textrm{ZF}} } \mathbf{P}^{-\frac{1}{2}} \mathbf{A}_n^{\textrm{T}} \tilde{ \mathbf{x} }_{:,n}.
    \label{eq:06}
\end{equation}
It follows that the ZF precoding presented in \cite{Dinis:2018} is given as 
\begin{equation}
    \tilde{ \mathbf{x} }_{:,n}^{\textrm{ZF}} = \beta_{\textrm{ZF}} \mathbf{A}_n^* ( \mathbf{A}_n^{\textrm{T}} \mathbf{A}_n^* )^{-1} \mathbf{ P }^{ \frac{1}{2}} \tilde{ \mathbf{s} }_{:,n}, 
    \label{eq:07}
\end{equation}
where $(\ )^*$ represents the complex conjugate operator.  In order to maintain the transmission power defined by $\textrm{tr} \left[ \mathbf{P} \right]$, $\beta_{\textrm{ZF}}$ must be set to the square root of $K$ divided by the gain associated with $\mathbf{A}_n^* ( \mathbf{A}_n^{\textrm{T}} \mathbf{A}_n^* )^{-1}$, namely
\begin{align}
	\beta_{\textrm{ZF}} &=\sqrt{ \frac{ K } { \mathbb{E} \{ \textrm{tr} [ ( \mathbf{A}_n^* ( \mathbf{A}_n^{\textrm{T}} \mathbf{A}_n^* )^{-1} )^{ \textrm{H} } \mathbf{A}_n^* ( \mathbf{A}_n^{\textrm{T}} \mathbf{A}_n^* )^{-1} ] \} } } \nonumber \\
	&= \sqrt{ \frac{ K } { \mathbb{E} \{ \textrm{tr} [ ( \mathbf{A}_n^{\textrm{T}} \mathbf{A}_n^* )^{-1} ] \} } } =  \sqrt{ \frac{ K } { \frac{ K } { M - K } } } \nonumber \\
	&= \sqrt{ M - K }.
	\label{eq:08}
\end{align}
The first part of the second expression of \eqref{eq:08} results from the fact that $ ( \mathbf{A}_n^{\textrm{T}} \mathbf{A}_n^* )^{-1}$ is Hermitian symmetric, and the second part follows from Lemma 2.10 of \cite{Verdu:Random_Matrix_Theory}.  The ZF precoding only requires a $K \times K$ matrix inversion, which is about the same computational complexity as the detector used in \cite{Kenney:Globecom:2020} for the UL.

%-------------------------------------------------------------------%
%-- Subsection IV.B: Regularized ZF Precoding with Matrix Invervse Reuse
%-------------------------------------------------------------------%
\subsection{Regularized ZF Precoding with Matrix Inverse Reuse} \label{Regularized ZF Precoding with Matrix Inverse Reuse}
An alternative approach to calculating the precoded vector is to solve \eqref{eq:06} for $\tilde{\mathbf{x}}_{:,n}$ in the following way (note that $\beta_{\textrm{ZF}}$ is replaced by $\beta_{\textrm{RZF}}$).  We start by multiplying both sides of \eqref{eq:06} by $\beta_{\textrm{RZF}} \mathbf{A}_n^* \mathbf{P}^{1/2}$.  The next step is to invert the $M$-dimensional square matrix $ \mathbf{A}_n^* \mathbf{A}_n^{\textrm{T}}$.  Since $\mathbf{A}_n^*$ is only of rank $K$ and $K < M$, the matrix is singular.  By adding a regularization factor to the matrix prior to inverting, we can obtain the following solution:
\begin{equation}
    {\tilde{\mathbf{x}}}_{:,n}^{\textrm{RZF}}= \beta_{\textrm{RZF}} \left(  \mathbf{A}_n^* \mathbf{A}_n^{\textrm{T}} + \sigma^2_w \mathbf{I}_M \right)^{-1}  \mathbf{A}_n^* \mathbf{ P }^{\frac{1}{2}} \tilde{ \mathbf{s} }_{:,n}, 
    \label{eq:09}
\end{equation}
where $\beta_{\textrm{RZF}}$ is a scale factor and $\sigma^2_w \mathbf{I}_M$ is chosen as the regularization term.  This regularization term selection will be shown to reuse the matrix inversion calculation performed for the UL.  To reduce the size of the matrix inverse, we next apply the matrix inversion lemma from \cite{Woodbury:1950}, which yields
\begin{equation}
	\tilde{\mathbf{x}}_{:,n}^{\textrm{RZF}} = \beta_{\textrm{RZF}} \mathbf{A}_n^* \left( \mathbf{A}_n^{\textrm{T}} \mathbf{A}_n^* + \sigma^2_w \mathbf{I}_K \right)^{-1} \mathbf{ P }^{\frac{1}{2}} \tilde{ \mathbf{s} }_{:,n}.
	\label{eq:10}
\end{equation}
Note that the RZF precoder defined in \eqref{eq:10} is based on a $K \times K$ matrix inversion, which is the complex conjugate of the matrix inversion used for the UL detector in \cite{Kenney:Globecom:2020}.  This result is highly desirable because reusing the matrix inverse that was calculated for the UL drastically reduces the computational complexity of the DL precoding.  

As in the ZF case, a transmission scale factor is used to maintain an average transmission power of $\textrm{tr} [ \mathbf{P} ]$.  Using the same form as \eqref{eq:08}, we have the following equation:
\begin{align}
	\beta_{\textrm{RZF}} &=\sqrt{ \frac{ K } { \mathbb{E} \{ \textrm{tr} [ ( \mathbf{A}_n^* ( \mathbf{A}_n^{\textrm{T}} \mathbf{A}_n^* + \sigma^2_w \mathbf{I}_K )^{-1} )^{ \textrm{H} } \mathbf{A}_n^* ( \mathbf{A}_n^{\textrm{T}} \mathbf{A}_n^* + \sigma^2_w \mathbf{I}_K )^{-1} ] \} } } \nonumber \\
	&= \frac{ 1 } { \sqrt{ \mathbb{E} \left\{ \lambda_{\beta} \right\} } },
	\label{eq:11}
\end{align}
where $\lambda_{\beta}$ is the average eigenvalue of the matrix shown in the first expression of \eqref{eq:11}.  The second expression in \eqref{eq:11} results from the fact that the trace of a matrix is equal to the sum of its eigenvalues \cite{Strang:1988}.  Using the value of $K$ from the numerator, the trace can be replaced by the average eigenvalue of the matrix.  The average is dropped, since the expectation is taken.  Since $\mathbf{A}_n^{\textrm{T}} \mathbf{A}_n^*$ is a Wishart matrix, its eigenvalues have the following probability distribution function (PDF) based on Theorem 2.17 of \cite{Verdu:Random_Matrix_Theory}:
\begin{equation}
	f_{ \lambda } (z) = \frac{1}{K} \sum_{ k=0 }^{ K-1 } \left( \frac{ k! \left( L_k^{ M-K } (z) \right)^2 }{ \left( k + M - K \right)! } \right) z^{ M-K } e^{-z}, \label{eq:12}
\end{equation}
where $ L_k^{ M-K } (z)$ is the Laguerre polynomial of order $k$.

After accounting for the differences between the matrix from \eqref{eq:11} and the Wishart matrix $\mathbf{A}_n^{\textrm{T}} \mathbf{A}_n^*$, the value of $\lambda_{\beta}$ is calculated as
\begin{equation}
	\lambda_{\beta} = \int_0^{ \infty} f_{\lambda} (z) \frac{ z }{ \left( z + \sigma^2_w \right)^2 } dz. \label{eq:13}
\end{equation}

There is no closed form solution to the integral above, but the $\lambda_{\beta}$ value can be calculated for a given value of $\sigma^2_w$ using numerical integration techniques.  The resulting value of $\beta_{\textrm{RZF}}$ for values of $K$ and $\sigma^2_w$ can be stored in a 2D look-up table at the base station.  An equivalent set of values is also needed at each UE to properly scale the received signal.

%-------------------------------------------------------------------%
%-- Subsection IV.C: UE Detection
%-------------------------------------------------------------------%
\subsection{UE Detection} \label{UE Detection}
As shown in \eqref{eq:05}, the precoded signal is multiplied by $ \mathbf{P}^{-\frac{1}{2}} \mathbf{A}_n^{\textrm{T}}$ as it traverses the channel.  In order to produce an unbiased estimate of the FD symbols, the received signal must be properly scaled at the UE.  The expression for the detected symbols is
\begin{align}
	 \hat{ \tilde{ \mathbf{s} } }_{:,n} &= \frac{ \alpha } { \beta_{\textrm{RZF}} } \left( \mathbf{P}^{-\frac{1}{2}} \mathbf{A}_n^{\textrm{T}} \tilde{\mathbf{x}}_{:,n}^{\textrm{RZF}} + \mathbf{w}_{:,n} \right) \nonumber \\
	&= \alpha \mathbf{P}^{-\frac{1}{2}} \mathbf{A}_n^{\textrm{T}} \mathbf{A}_n^* ( \mathbf{A}_n^{\textrm{T}} \mathbf{A}_n^* + \sigma^2_w \mathbf{I}_K )^{-1} \mathbf{ P }^{\frac{1}{2}} \tilde{ \mathbf{s} }_{:,n} + \frac{ \alpha \tilde{ \mathbf{w} }_{:,n} }{ \beta_{\textrm{RZF}} },
	\label{eq:14}
\end{align}
where the scalar $\alpha$ is included to compensate for the scaling that results from the matrix $\mathbf{A}_n^{\textrm{T}} \mathbf{A}_n^* ( \mathbf{A}_n^{\textrm{T}} \mathbf{A}_n^* + \sigma^2_w \mathbf{I}_K )^{-1}$.  The unbiased condition is met if $\alpha$ is set such that
\begin{equation}
	\tilde{ \mathbf{s} }_{:,n} = \alpha \mathbb{E} \left\{ \mathbf{P}^{-\frac{1}{2}} \mathbf{A}_n^{\textrm{T}} \mathbf{A}_n^* ( \mathbf{A}_n^{\textrm{T}} \mathbf{A}_n^* + \sigma^2_w \mathbf{I}_K )^{-1} \mathbf{ P }^{\frac{1}{2}} \right\} \tilde{ \mathbf{s} }_{:,n}. \label{eq:15}
\end{equation}

The expectation in \eqref{eq:15} must be set to $\frac{ 1 }{ \alpha } \mathbf{I}_K$ in order for \eqref{eq:15} to be satisfied.  This result can be simplified as
\begin{align}
	\frac{ 1 }{ \alpha } \mathbf{I}_K &= \mathbb{E} \left\{ \mathbf{P}^{-\frac{1}{2}} \mathbf{A}_n^{\textrm{T}} \mathbf{A}_n^* ( \mathbf{A}_n^{\textrm{T}} \mathbf{A}_n^* + \sigma^2_w \mathbf{I}_K )^{-1} \mathbf{ P }^{\frac{1}{2}} \right\} \nonumber \\
	&= \mathbb{E} \left\{ \mathbf{A}_n^{\textrm{T}} \mathbf{A}_n^* ( \mathbf{A}_n^{\textrm{T}} \mathbf{A}_n^* + \sigma^2_w \mathbf{I}_K )^{-1} \right\} \nonumber \\
%	&= \mathbb{E} \left\{ \mathbf{I}_K - \sigma_w^2 ( \mathbf{A}_n^{\textrm{T}} \mathbf{A}_n^* + \sigma^2_w \mathbf{I}_K )^{-1} \right\} \nonumber \\
	&= \mathbf{I}_K - \sigma_w^2  \mathbb{E} \left\{ ( \mathbf{A}_n^{\textrm{T}} \mathbf{A}_n^* + \sigma^2_w \mathbf{I}_K )^{-1} \right\}. \label{eq:16}
%	&= \mathbf{I}_K - \sigma_w^2  \mathbb{E} \left\{ \frac{ \textrm{tr} \left[ ( \mathbf{A}_n^{\textrm{T}} \mathbf{A}_n^* + \sigma^2_w \mathbf{I}_K )^{-1} \right] }{ K } \right\} \mathbf{I}_K  \nonumber \\
%	&= \mathbf{I}_K - \sigma_w^2  \lambda_{\alpha} \mathbf{I}_K, \label{eq:16}
\end{align}
Since $\mathbf{A}_n^{\textrm{T}} \mathbf{A}_n^*$ is a Wishart matrix, the expected value of the off-diagonal values of $( \mathbf{A}_n^{\textrm{T}} \mathbf{A}_n^* + \sigma^2_w \mathbf{I}_K )^{-1}$ is zero.  The expected value of the diagonal entries is a constant.  Hence, \eqref{eq:16} can be further reduced to
\begin{equation}
	\frac{ 1 }{ \alpha } \mathbf{I}_K = \mathbf{I}_K - \sigma_w^2  \lambda_{\alpha} \mathbf{I}_K, \label{eq:16a}
\end{equation}
where $\lambda_{\alpha}$ is equal to $\mathbb{E} \left\{ \textrm{tr} \left[ ( \mathbf{A}_n^{\textrm{T}} \mathbf{A}_n^* + \sigma^2_w \mathbf{I}_K )^{-1} \right] \right\} / K$, which is equal to the average eigenvalue of $( \mathbf{A}_n^{\textrm{T}} \mathbf{A}_n^* + \sigma^2_w \mathbf{I}_K )^{-1}$.

Although the expression for $\lambda_{\alpha}$ has no closed form solution, it can be calculated by computing the integral of the probability distribution function (PDF).  We note that $\mathbf{A}_n^{\textrm{T}} \mathbf{A}_n^*$ is a Wishart matrix, which has the PDF shown in \eqref{eq:12} for its eigenvalues.  Using the PDF in \eqref{eq:12}, we can express $\lambda_{\alpha}$ as
\begin{equation}
	\lambda_{\alpha} = \int_0^{ \infty} \frac{ f_{\lambda} (z) }{ z + \sigma^2_w } dz. \label{eq:17}
\end{equation}
The value in \eqref{eq:17}, can be readily calculated for a given value of $\sigma^2_w$, $K$, and $M$ using numerical integration techniques.  Like the $\beta_{\textrm{RZF}}$ scalar, the values for $\alpha$ can be pre-computed and stored in a 2D table based on $K$ and $\sigma_w^2$.  The value of $\lambda_{\alpha}$ is simulated by taking $1/K$ times the average value of the trace of several random instantiations of a Wishart matrix, $\mathbf{W} = \mathbf{A}^{\textrm{T}} \mathbf{A}^*$, where the elements of $\mathbf{A}$ are i.i.d. complex Gaussian, zero-mean, and unit variance.  For example, simulated results with $10^4$ iterations match very closely with the results obtained through numerical integration.

Once $\lambda_{\alpha}$ is calculated, it can be used to compute $\alpha$ based on \eqref{eq:16a} as follows:
\begin{equation}
	\alpha = \frac{ 1 }{ 1 - \sigma^2_w \lambda_{\alpha} }. \label{eq:18}
\end{equation} 

Asymptotes for the value of $\alpha$ are easily attained by examining \eqref{eq:14} in the low and high SNR regimes.  In the low SNR case, $\sigma^2_w$ dominates the inverse in \eqref{eq:14} such that it converges to $\frac{1}{ \sigma^2_w } \mathbf{I}_K$.  Since the expected value of each diagonal element of $\mathbf{A}_n^{\textrm{H}} \mathbf{A}_n $ is $M$, the asymptotic value for $\alpha$ in the low SNR regime is $\frac{ \sigma^2_w }{ M }$.  In the high SNR case, the matrix inverse converges to $( \mathbf{A}_n^{\textrm{H}} \mathbf{A}_n )^{-1}$, resulting in a unity scaling for each symbol.  As a result, the asymptotic value for $\alpha$ in the high SNR regime is unity.

%-------------------------------------------------------------------%
%-- Subsection IV.C: Power Optimization
%-------------------------------------------------------------------%
\subsection{Power Optimization} \label{Power Optimization}
%
%Since the matrix inverse in \eqref{eq:14} is reused from the UL detection, where the signals from all UEs are presented at the base station at equal power due to power control, the performance in \eqref{eq:14} is optimized for case where the diagonals of $\mathbf{P}$ are equal.  
When the large-scale path loss varies greatly for different UEs, the UEs that are close to the BS will see higher levels of MUI than the UEs that are at the edge of the cell.  Based on \eqref{eq:14}, the output SINR for user $k$ is
\begin{equation}
	\gamma_k = \frac{ 1 }{  \alpha^2 \sigma^2_{\!\textrm{od}} \frac{ \left( p_{\textrm{total} } - p_k \right) }{ p_k }+ \frac{ \alpha^2 }{ \beta^2_{\textrm{RZF}} } \sigma_w^2 }, \label{eq:19}
\end{equation}
where $\sigma^2_{\!\textrm{od}}$ is the variance for all of the off-diagonal elements of $\mathbf{A}_n^{\textrm{T}} \mathbf{A}_n^* ( \mathbf{A}_n^{\textrm{T}} \mathbf{A}_n^* + \sigma^2_w \mathbf{I}_K )^{-1}$, and $p_{\textrm{total} } = \sum_{k=1}^K p_k$.  The numerator of \eqref{eq:19} is unity because the scaling previously described results in an unbiased estimate.  The first term in the denominator is the MUI, which is based on the power of the other users (i.e., $p_{\textrm{total} } - p_k$).  The MUI is divided by $p_k$ due to the large-scale path loss for user $k$.  The second term in the denominator is the receiver noise with its appropriate scaling.  For the case of equal power with equal large-scale path loss, we find that each UE achieves an output SINR of
\begin{equation}
	\gamma_{\textrm{eq}} = \frac{ 1 }{ \alpha^2 \sigma^2_{\!\textrm{od}} \left( K - 1 \right) + \frac{ \alpha^2 }{ \beta^2_{\textrm{RZF}} } \sigma_w^2 }. \label{eq:20}
\end{equation}

In order to achieve the same SINR for each UE when large-scale path loss is not equal, the amplitude scaling of $\mathbf{P}^{1/2}$ in \eqref{eq:10} and \eqref{eq:14} is replaced by the diagonal matrix $\mathbf{Q}^{1/2}$, where $\mathbf{Q}$ has diagonal elements of $q_k$.  In order to not introduce any bias from the new scaling, each UE multiplies its received signal by an additional scale factor of $\sqrt{p_k / q_k}$.  Consequently, each UE must have knowledge of the channel statistics as well as the amplitude scaling factor applied by the BS.  The estimates of the FD symbols in \eqref{eq:14} now change to
\begin{equation}
	 \hat{ \tilde{ \mathbf{s} } }_{:,n} = \alpha \mathbf{ Q }^{-\frac{1}{2}} \mathbf{A}_n^{\textrm{T}} \mathbf{A}_n^* ( \mathbf{A}_n^{\textrm{T}} \mathbf{A}_n^* + \sigma^2_w \mathbf{I}_K )^{-1} \mathbf{ Q }^{\frac{1}{2}} \tilde{ \mathbf{s} }_{:,n} + \frac{ \alpha \mathbf{ Q }^{- \frac{1}{2}} \mathbf{ P }^{\frac{1}{2}} \tilde{ \mathbf{w} }_{:,n} }{ \beta_{\textrm{RZF}} }. \label{eq:21}
\end{equation}
The values for $q_k$ are set to achieve the same SINR for each user, $\gamma$, while abiding by the power constraint of $\sum_{k=1}^K q_k = p_{\textrm{total} }$.  The output SINR for each user with power optimization is expressed as
\begin{equation}
	\gamma = \frac{ 1 }{ \alpha^2 \sigma^2_{\!\textrm{od}} \frac{ \left( p_{\textrm{total} } - q_k \right) }{ q_k } + \frac{ \alpha^2 p_k }{ \beta^2_{\textrm{RZF}} q_k } \sigma_w^2 }. \label{eq:22}
\end{equation}

An expression for the achieved value of $\gamma$ is obtained by rearranging \eqref{eq:22} and summing over all $K$ UEs, which yields
\begin{align}
	\sum_{k=1}^K \gamma \left( \alpha^2 \sigma^2_{\!\textrm{od}} \left( p_{\textrm{total} } - q_k \right) + \frac{ \alpha^2 }{ \beta^2_{\textrm{RZF}} } p_k \sigma_w^2 \right) &= \sum_{k=1}^K q_k \nonumber \\
	\gamma \left( \alpha^2 \sigma^2_{\!\textrm{od}} \left( K p_{\textrm{total} } -  p_{\textrm{total} } \right) + \frac{ \alpha^2 }{ \beta^2_{\textrm{RZF}} }  p_{\textrm{total} } \sigma_w^2 \right) &=  p_{\textrm{total} } \nonumber \\
	\gamma = \frac{ 1 }{ \alpha^2 \sigma^2_{\!\textrm{od}} \left( K - 1 \right) + \frac{ \alpha^2 }{ \beta^2_{\textrm{RZF}} } \sigma_w^2 } &= \gamma_{\textrm{eq}}. \label{eq:23}
\end{align}
The result in \eqref{eq:23} shows that the power optimization produces the same performance that is achieved in the case of equal power with equal large-scale path loss.  Equating the expressions in \eqref{eq:22} and \eqref{eq:23} and solving for $q_k$ we arrive at
\begin{equation}
	q_k = \frac{ \sigma^2_{\!\textrm{od}} p_{\textrm{total}} + p_k \frac{ \sigma_w^2  }{ \beta^2_{\textrm{RZF}} } }{ K \sigma^2_{\!\textrm{od}} + \frac{ \sigma_w^2  }{ \beta^2_{\textrm{RZF}} } }.
\end{equation}

%-------------------------------------------------------------------%
%-- Section IV: Complexity Comparison
%-------------------------------------------------------------------%
\section{Complexity Comparison} \label{Complexity_Comparison}
\begin{figure}[!t]
\centering
\includegraphics[width=3.6in, clip=true, trim=4cm 8.5cm 4cm 9cm ]{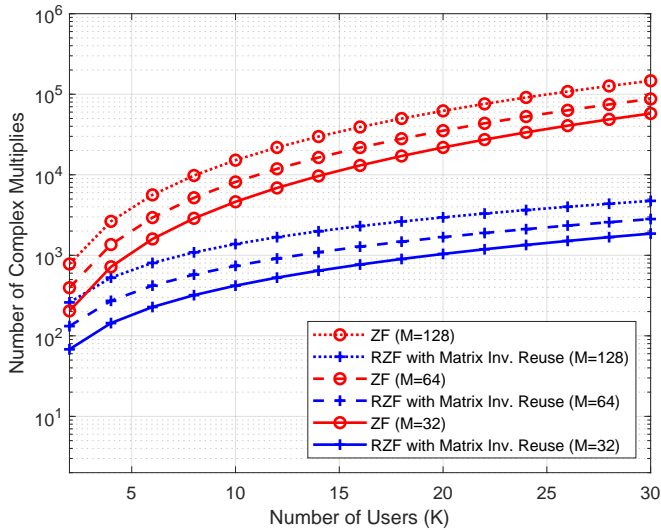}
\caption{The number of complex multiplies per bin is plotted versus the number of users for the two precoding implementations and for different numbers of BS antennas (i.e., $M=32$, $64$, and $128$).  The ZF precoder, here, follows the formulation presented in \cite{Dinis:2018}.  The RZF approach that reuses the matrix inversion has significant savings, especially as $K$ grows large.  }
\label{complexity_comp_DL}
\end{figure}

By reusing the matrix inversion from the UL detector, the RZF precoder is much more computationally efficient than the ZF precoder of \cite{Dinis:2018}.  A comparison in terms of the number of complex multiplies is presented here.  Since both algorithms require conversion to and from the frequency domain, this operation is not considered in the comparison.  Likewise, since the amplitude scaling by $\mathbf{P}^{1/2}$ or $\mathbf{Q}^{1/2}$ is the same, it is also excluded even though its contribution is insignificant.  As before, we assign $K^3$ complex multiplies for a $K \times K$ matrix inversion.  The number of complex multiples involved in a matrix (or vector) multiplication is the product of the outer dimensions times the inner dimension.

The following operations are required for each bin of the ZF precoding in \eqref{eq:07}, resulting in a total of $K^2 + KM + K^3 + K^2 M$ complex multiplies per bin:
\begin{description}
	\item[$K^2 M$] Multiply $(K \times M)$ matrix by $(M \times K)$ matrix
	\item[$K^3$] Invert the resulting $(K \times K)$ matrix
	\item[$K^2$] Multiply $(K \times K)$ matrix by $(K \times 1)$ vector
	\item[$KM$] Multiply $(M \times K)$ matrix by $(K \times 1)$ vector
\end{description}

To reuse the matrix inversion from the UL detector, the complex conjugate is taken, which does not require any complex multiplies.  Thus, the computational complexity of the RZF precoder in \eqref{eq:10} is reduced to the a total of $K^2 + KM$ multiplies per bin:
\begin{description}
	\item[$K^2$] Multiply $(K \times K)$ matrix by $(K \times 1)$ vector
	\item[$KM$] Multiply $(M \times K)$ matrix by $(K \times 1)$ vector
\end{description}
Fig. \ref{complexity_comp_DL} shows a comparison of the ZF and the RZF precoder.

%-------------------------------------------------------------------%
%-- Section V: Performance Analysis
%-------------------------------------------------------------------%
\section{Performance Analysis} \label{Performance_Analysis}
The DL performance of the RZF precoder is very similar to the UL detector performance reported in \cite{Kenney:Globecom:2020}.  The estimated symbol expression in \eqref{eq:14} will be analyzed for the low-SNR and the high-SNR cases, using the expressions for $\beta_{\textrm{RZF}}$ presented previously for each case.

%-------------------------------------------------------------------%
%-- Subsection V.A: Low SNR Operation
%-------------------------------------------------------------------%
\subsection{Low SNR Operation}
When the noise variance dominates the expression for the matrix inverse in \eqref{eq:14}, the matrix inverse will converge to the identity matrix scaled by the inverse of the noise variance.  The value of $\alpha$ converges to $\sigma^2_w / M$, and $\beta_{\textrm{RZF}} = \sigma^2_w / \sqrt{M}$ in the low-SNR regime as shown below:
\begin{align}
	\beta_{\textrm{RZF,Low-SNR}} &=\sqrt{ \frac{ K } { \frac{1}{ \sigma^4_w } \mathbb{E} \left\{ \textrm{tr} \left[ \mathbf{A}_n^{ \textrm{T} } \mathbf{A}_n^*  \right] \right\} } } \nonumber \\
	&= \sigma^2_w \sqrt{ \frac{ K }{ MK } } \nonumber \\
	&= \frac{ \sigma^2_w }{ \sqrt{ M } },
	\label{eq:24}
\end{align}
where the second expression results from Lemma 2.9 of \cite{Verdu:Random_Matrix_Theory}.

The resulting expression for the Low-SNR DL symbol estimates in the FD is
\begin{equation}
	\hat{\tilde{\mathbf{s}}}_{:,n}^{\textrm{Low-SNR,DL}} = \frac{1}{M} \mathbf{P}^{-\frac{1}{2}} \mathbf{A}_n^{\textrm{T}} \mathbf{A}_n^* \mathbf{P}^{\frac{1}{2}} \tilde{ \mathbf{s} }_{:,n} + \frac{ \tilde{ \mathbf{w} }_{:,n} }{ \sqrt{M}}.
	\label{eq:25}
\end{equation}
The effect of the $\mathbf{P}$ matrices were discussed at the end of Section \ref{Power Optimization}.  With the aforementioned method of mitigating the effect of large disparities in the path loss between users, we can set $\mathbf{P} = \mathbf{I}_K$ for the performance analysis.  With this convention, we see that the expected value of the diagonals of $\mathbf{A}_n^{\textrm{T}} \mathbf{A}_n^*$ are equal to $M$, which results in an unbiased estimate.  The noise variance of $\sigma^2_w$ is divided by $M$.  The resulting SINR is equal to $M/ \sigma^2_w$ (i.e., performance gain of $M$), since the noise dominates the MUI caused by the off-diagonal elements of $\mathbf{A}_n^{\textrm{T}} \mathbf{A}_n^*$ in the low-SNR regime.  This is an identical result to the low SNR UL case in \cite{Kenney:Globecom:2020}.

%-------------------------------------------------------------------%
%-- Subsection V.B: High SNR Operation
%-------------------------------------------------------------------%
\subsection{High SNR Operation}
At high SNR the noise variance becomes vanishingly small.  As a result, the matrix inverse in \eqref{eq:14} reduces to $(\mathbf{A}_n^{\textrm{T}} \mathbf{A}_n^*)^{-1}$, which cancels the $\mathbf{A}_n^{\textrm{T}} \mathbf{A}_n^*$ matrix.  Then the $\mathbf{P}^{-\frac{1}{2}}$ term cancels with the $\mathbf{P}^{\frac{1}{2}}$ term.  At high SNR, $\alpha$ converges to unity, and $\beta_{\textrm{RZF}}$ converges to $\beta_{\textrm{ZF}}$ in \eqref{eq:08}.  The expression for the symbol estimates in the FD is simplified to  
\begin{equation}
	\hat{\tilde{\mathbf{s}}}_{:,n}^{\textrm{High-SNR,DL}} = \tilde{\mathbf{s}}_{:,n}  + \frac{ \tilde{ \mathbf{ w } }_{:,n} } { \sqrt{ M - K } }.
	\label{eq:26}
\end{equation}

Based on \eqref{eq:26}, the high-SNR performance has a performance gain of $M-K$ due to the scaling of the noise.  Unlike the low SNR case, the near-far effect does not contribute to the interference at high SNR.  This is an identical result to the high SNR UL case in \cite{Kenney:Globecom:2020}.

%-------------------------------------------------------------------%
%-- Subsection VI: Simulation Results
%-------------------------------------------------------------------%
\section{Simulation Results} \label{Simulation_Results}
\begin{figure}[!t]
    \centering
    \includegraphics[width=3.6in, clip=true, trim=4cm 8.5cm 4cm 9cm ]{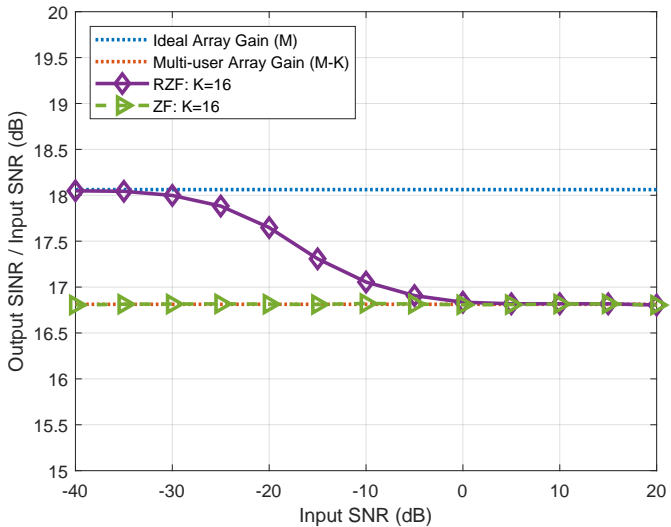}
    \caption{Ratio of output SINR to input SNR (i.e., performance gain) is plotted versus the input SNR for the DL scenario where $M=64$ and $K=16$.  Both RZF and ZF performance gains match the values in the performance analysis.  RZF has a sizeable advantage over the ZF precoder at low input SNR values.  }
    \label{DL_performance}
\end{figure}

A single-cell DL scenario was simulated to show the performance of the FD multi-user precoding techniques detailed in this section.  During each frame, $N=2048$ samples are transmitted.  The BS transmits to $K$ UEs simultaneously.  In order to reach each UE with the same power, the BS must transmit more power to more distant UEs with higher large-scale path losses.  In this simulation, the excess path loss for each UE is distributed between $0$ dB and $20$ dB.  Because of the additional MUI created by the large variation in transmit power of the different signal components, the power optimization method presented in Section \ref{Power Optimization} was simulated.  The resulting performance matches the case where the large-scale path losses are equal.

Each channel impulse response between BS antenna $m$ and UE $k$ is randomly selected with an exponential power delay profile and a roll-off factor of 25 samples.  The length of the channel impulse response, $L_h$, is set to 130 samples.  It should be noted that the performance of this algorithm is independent of the value of $L_h$ as long as $L_h \leq L_{\textrm{CP}} - 1$, where $L_{\textrm{CP}}$ is the length of the CP.  The average power of each channel for each user is set to unity, but the individual power values are uniformly distributed between $0.1$ and $2.0$.  Slight modifications to the scaling are made to maintain the unity average power constraint after the individual power values are chosen.  These results assume perfect CSI is available at the BS.

The main objective of the simulation is to measure the resulting SINR after being detected at the receiver and then averaged across all UEs.  Because the BS has $M$ antennas, an ideal antenna gain of $M$ would be expected for a single-user case (i.e., the output SINR could be as high as $M$ times the input SNR), which is also the predicted performance at low input SNR.  Based on the value of $\beta_{\textrm{ZF}}$, the ZF precoder performance gain is expected to be $M-K$ over all input SNR values.  This is also the predicted performance for the RZF precoder case at high input SNR.  Fig. \ref{DL_performance} shows the results of the simulation.

%----------------------------%
%-- Section VII: Conclusion --%
%----------------------------%
\section{Conclusion} \label{Conclusion}
This paper presented a RZF precoding solution that reuses the matrix inverses calculated as part of the UL detection as presented in \cite{Kenney:Globecom:2020}.  Derivations for the scale factors necessary to achieve the transmit power target and to obtain unbiased symbol estimates at the UE were provided.  In addition, a transmit power optimization was presented that compensates for MUI effects that reduce the average SINR at the UEs when large variations in the large-scale path loss exist.  Because of the reuse of the matrix inverse for each processing bin, it was shown that the RZF precoding solution requires drastically fewer complex multiplies than the ZF precoder.  The performance analysis showed that the achievable output SINR from the RZF precoder is greater than that of the ZF precoder for moderate and low input SNR values.  It was also shown that the RZF precoder performance converged to that of the ZF precoder for high SNR values.  The analysis was verified via simulation of a single-cell massive MIMO scenario with $64$ BS antennas and $16$ UEs.  With the low computational complexity and superior performance at moderate input SNR values, we conclude that RZF is the preferred approach to massive MIMO DL precoding for SCM waveforms with a CP.

%-- Bibliography --%

\end{document}